\documentclass[conference, 10pt]{IEEEtran}
\IEEEoverridecommandlockouts

\usepackage{cite}
\usepackage{amsmath,amssymb,amsfonts}
\usepackage{algorithmic}
\usepackage{graphicx}
\usepackage{textcomp}
\usepackage[dvipsnames]{xcolor}

\usepackage{hyperref}

\usepackage{comment}
\usepackage{wrapfig}
\usepackage{pgfplots}
\pgfplotsset{compat=1.18}

\usepackage{tikz}
\usetikzlibrary{fit}
\usetikzlibrary{shapes,arrows}
\usetikzlibrary{positioning}
\usetikzlibrary{shapes.geometric}
\usetikzlibrary{chains, decorations.markings, shadows, shapes.arrows, shapes, fit}
\usepackage{circuitikz}
\usepackage{multirow}
\usepackage{overpic}
\usetikzlibrary{positioning, fit}

\def\BibTeX{{\rm B\kern-.05em{\sc i\kern-.025em b}\kern-.08em
    T\kern-.1667em\lower.7ex\hbox{E}\kern-.125emX}}
\begin{document}

\title{Radar Detection through Rectified Flow Matching
\thanks{Part of this work was supported by ANR ASTRID Neptune 3 (ANR-23-ASM2-0009)}
}

\author{
\IEEEauthorblockN{
P. Meena\IEEEauthorrefmark{1},
Y. A. Rouzoumka\IEEEauthorrefmark{2}\IEEEauthorrefmark{3},
J. Pinsolle\IEEEauthorrefmark{2},
C. Ren\IEEEauthorrefmark{2},
M. N. El Korso\IEEEauthorrefmark{1},
J.-P. Ovarlez\IEEEauthorrefmark{2}\IEEEauthorrefmark{3}
}
\IEEEauthorblockA{\IEEEauthorrefmark{1}
\textit{L2S, CentraleSupélec, Université Paris-Saclay, 91190 Gif-sur-Yvette, France} \\
}
\IEEEauthorblockA{\IEEEauthorrefmark{2}
\textit{SONDRA, CentraleSupélec, Université Paris-Saclay, 91190 Gif-sur-Yvette, France} \\
}
\IEEEauthorblockA{\IEEEauthorrefmark{3}
\textit{DEMR, ONERA, Université Paris-Saclay, 91120 Palaiseau, France} \\
}
}

\maketitle

\begin{abstract}
Radar target detection in the presence of a mixture of non-Gaussian clutter and white thermal noise is a challenging problem. This paper proposes a Rectified Flow Matching-based method for radar detection, termed D-RFM. Unlike existing detectors, D-RFM learns a mapping from a standard Gaussian distribution to radar observations by capturing the underlying velocity field. Detection is then performed by inverse mapping test samples into the latent Gaussian space using the learned velocity field, with targets identified as deviations from the learned distribution. Experimental results demonstrate the efficacy of the proposed method under both Gaussian and non-Gaussian clutter plus additive white Gaussian noise, highlighting its accuracy, robustness, and computational efficiency. 

\end{abstract}


\section{Introduction}
\label{sec:intro}

Radar systems have gained increasing importance in a wide range of civilian and military applications due to their robustness, long-range sensing capability, and ability to operate under adverse weather and illumination conditions \cite{richards2005fundamentals}. Applications include surveillance, navigation, environmental monitoring, and weather forecasting. The widespread deployment of radar systems has resulted in extensive research focused on developing advanced radar signal processing techniques.

Target detection is a fundamental and widely studied problem in radar signal processing, involving the identification of the presence of the target in a noisy environment. Accurate detection is challenging due to weak target signals at low signal-to-noise ratio (SNR), non-Gaussian and heterogeneous clutter, and environmental variations \cite{de2016modern}. Traditional target detection methods, including the Matched Filter (MF), Normalized Matched Filter (NMF) \cite{scharf2002matched}, Adaptive Matched Filter (AMF) \cite{fuhrmann1992cfar}, Kelly detector \cite{kelly2007adaptive}, and Adaptive Normalized Matched Filter (ANMF) \cite{kraut2001adaptive}, rely on assumptions of Gaussian and homogeneous clutter, and often perform suboptimally in low SNR, non-Gaussian, or heterogeneous clutter environments. Moreover, conventional detectors primarily exploit low-level features, which are sensitive to noise and clutter variations.

Compared with traditional methods, deep learning (DL)–based methods \cite{goodfellow2016deep, zhang2024ifcmd} have demonstrated superior target detection performance by learning high-level feature representations from data. By leveraging high-level features, DL models such as Convolutional Neural Networks (CNNs) \cite{yavuz2021radar} and Recurrent Neural Networks (RNNs) \cite{sehgal2019automatic} can mitigate the limitations of conventional detectors. Furthermore, DL-based methods reduce false alarms, which improves their adaptability to diverse radar conditions. Existing DL methods can be broadly classified into supervised and unsupervised \cite{goodfellow2016deep}. Supervised methods formulate radar target detection as a classification problem and require accurately labeled training data, whereas unsupervised methods formulate detection as an anomaly detection problem, in which the model learns a feature-space representation of clutter (target-absent) samples and identifies target-present samples as outliers. However, the performance of DL-based methods is highly dependent on the availability of large-scale training data. In addition, these methods often exhibit degraded performance for small or slow-moving targets and suffer from limited generalization capability, which restricts their applicability in dynamic environments.

Recent works have employed deep generative models, such as Variational Autoencoders (VAEs) \cite{VAEalexis, rouzoumka2025complex} and Generative Adversarial Networks (GANs) \cite{zhao2024gan}, to learn the distribution of radar data (e.g., clutter and target). By capturing the underlying characteristics of radar signals, generative model–based methods formulate target detection as an anomaly detection problem, where anomalies are identified as deviations from the learned clutter distribution. Although these methods can improve detection performance, they often struggle to detect weak targets that exhibit subtle deviations from clutter, particularly under low SNR. Furthermore, generative models can be computationally expensive, require large-scale training data to accurately capture complex radar signal distributions, and often suffer from slow inference.

To address the above limitations, flow matching (FM) \cite{liu2022flow} provides a more efficient approach for modeling complex data distributions by learning a velocity field that maps a simple prior (e.g., a Gaussian) to the target distribution. This allows for accurate modeling of high-dimensional data with lower computational cost and faster inference compared to traditional generative models. Although FM has been successfully applied in image generation and translation \cite{liu2022flow}, computational biology \cite{li2025flow}, and radar signal processing \cite{mottier2024deinterleaving}, its application in radar target detection remains unexplored. To the best of our knowledge, this work is the first to leverage rectified flow matching (RFM) for radar target detection, enabling robust identification of weak targets in both Gaussian and non-Gaussian clutter environments while maintaining computational efficiency.
The main contributions of this paper are:
\noindent (i) A RFM-based framework for radar detection, termed D-RFM, that learns a deterministic velocity field (i.e., flow) mapping from Gaussian noise to the radar clutter distribution.
(ii) An unsupervised anomaly scoring mechanism based on the learned flow that enables reliable detection of low-SNR targets in dynamic radar environments.
(iii) A computationally efficient generative approach that employs deterministic flow with straight-line interpolation to provide robustness under non-Gaussian clutter and improved weak-target detection performance. 
    

\section{Statistical Model}

In radar detection, the objective is to decide between two hypotheses based on the received signal. Let the received signal be a complex vector $\mathbf{y}\in\mathbb{C}^N$. Then, the radar detection problem can be formulated as a binary hypothesis test:
\begin{equation}
\begin{cases} 
\mathcal{H}_0: \mathbf{y} = \mathbf{c} + \mathbf{n}, \\[1mm]
\mathcal{H}_1: \mathbf{y} = \alpha \, \mathbf{p} + \mathbf{c} + \mathbf{n},
\end{cases}
\label{eq:bihyp}
\end{equation}
where $\alpha \in \mathbb{C}$ denotes the unknown complex amplitude of the target, $\mathbf{p}\in \mathbb{C}^N$ is the steering vector, and $\mathbf{c}\in \mathbb{C}^N$ and $\mathbf{n}\sim \mathcal{CN}(\mathbf{0}, \sigma^2 \, \mathbf{I})$ represent clutter and the additive white Gaussian noise (AWGN), respectively.

In the homogeneous clutter case, $\mathbf{c}$ is modeled as a complex circular Gaussian vector, $ \mathbf{c} \sim \mathcal{CN}(\mathbf{0}, \mathbf{\Sigma}_c)$. In contrast, heterogeneous clutter is described using a compound Gaussian model, $\mathbf{c} = \sqrt{\delta} \, \mathbf{g}$, where $ \mathbf{g} \sim \mathcal{CN}(\mathbf{0}, \mathbf{\Sigma}_c)$ and, conditionally on the texture $\delta$, $\mathbf{c} \,|\, \delta \sim \mathcal{CN}(\mathbf{0}, \delta \,\mathbf{\Sigma}_c)$. Here, $\delta$ captures the power fluctuations across radar cells and is assumed to satisfy $\mathbb{E}[\delta] = 1$ for simplicity.

The SNR under hypothesis $\mathcal{H}_1$, after applying a whitening transformation, is defined as $\mathrm{SNR} = |\alpha|^2 \,\mathbf{p}^H \mathbf{\Sigma}^{-1} \mathbf{p}$. Here, $\mathbf{\Sigma} = \mathbf{\Sigma}_c + \sigma^2 \,\mathbf{I}$ denotes the covariance matrix of clutter-plus-noise. Throughout this paper, the power ratio between clutter and thermal noise used is set to $r = \mathrm{Tr}(\mathbf{\Sigma}_c)/(N \sigma^2) = 1$.

Numerous detectors have been proposed in the literature for radar target detection. 
Detection strategies depend on the knowledge of the noise statistics and environmental conditions. In a thermal noise-free environment, detection strategies have been developed for both homogeneous and partially homogeneous cases, as well as their adaptive counterparts.

In a homogeneous environment, when the noise covariance matrix $\mathbf{\Sigma}$ is known and uniform across observation space, the Matched Filter (MF) provides the optimal detection as:
\begin{equation}
    \Lambda_{\mathrm{MF}}(\mathbf{y}) = \frac{|\mathbf{p}^H \mathbf{\Sigma}^{-1} \mathbf{y}|^2}{\mathbf{p}^H \mathbf{\Sigma}^{-1} \mathbf{p}} 
    \;\underset{H_0}{\overset{H_1}{\gtrless}} \lambda \, ,
\label{eq:MF}
\end{equation} 
where $\lambda$ denotes the detection threshold.

In environments where clutter and thermal noise share a common covariance up to an unknown scaling factor, the Normalized Matched Filter (NMF) \cite{scharf2002matched} is commonly used due to its invariance to this unknown scale, and the detection is performed as:
\begin{equation}
    \Lambda_{\mathrm{NMF}}(\mathbf{y}) = \frac{|\mathbf{p}^H \mathbf{\Sigma}^{-1} \mathbf{y}|^2}
    {(\mathbf{p}^H \mathbf{\Sigma}^{-1} \mathbf{p})(\mathbf{y}^H \mathbf{\Sigma}^{-1} \mathbf{y})} 
    \;\underset{H_0}{\overset{H_1}{\gtrless}} \lambda\, .
\label{eq:NMF}
\end{equation}

When the noise covariance matrix $\mathbf{\Sigma}$ is unknown, but the noise remains Gaussian, adaptive detectors are employed. The adaptive detector estimates the covariance from $K$ secondary (clutter-only) observations $\{\mathbf{z}_k\}_{k=1}^K$ using the Sample Covariance Matrix (SCM):
\begin{equation}
    \hat{\mathbf{\Sigma}}_{\mathrm{SCM}} = \frac{1}{K} \sum_{k=1}^{K} \mathbf{z}_k \,\mathbf{z}_k^H\, .
\label{eq:SCM}
\end{equation}

By replacing $\mathbf{\Sigma}$ with $\hat{\mathbf{\Sigma}}_{\mathrm{SCM}}$ in the MF \eqref{eq:MF} and NMF \eqref{eq:NMF}, the resulting detectors are the Adaptive Matched Filter (AMF-SCM) \cite{fuhrmann1992cfar} and the Adaptive Normalized Matched Filter (ANMF-SCM) \cite{kraut2001adaptive}, respectively. 

Although adaptive detectors improve detection performance over classical MF and NMF, their performance degrades significantly in the presence of compound Gaussian clutter. To address this, the Tyler Adaptive Normalized Matched Filter (ANMF-FP) has been proposed \cite{de2016modern, tyler1987}, where the covariance matrix is estimated using the robust Tyler estimator based on the secondary observations $\{\mathbf{z}_k\}$:
\begin{equation}
    \hat{\mathbf{\Sigma}}_{\mathrm{FP}} = \frac{N}{K} \sum_{k=1}^{K} \frac{\mathbf{z}_k \,\mathbf{z}_k^H}{\mathbf{z}_k^H \,\hat{\mathbf{\Sigma}}_{\mathrm{FP}}^{-1} \,\mathbf{z}_k}\, .
    \label{eq:ANMF_FP}
\end{equation}
 
Although ANMF-FP performs well in compound Gaussian clutter environments, its performance degrades in the presence of additive thermal noise. In the case of the Gaussian clutter plus thermal noise, classical detectors (e.g., MF \cite{scharf2002matched}, AMF-SCM \cite{fuhrmann1992cfar}, and ANMF-SCM \cite{kraut2001adaptive}) remain valid, since the sum of Gaussian-distributed noise components is still Gaussian. However, when the clutter deviates from the Gaussian assumption, the limitations of these classical methods become evident. Kernel-based methods such as Support Vector Data Description (SVDD) \cite{svdd} have recently been adapted to target detection \cite{svdd_detection}. They provide a non-parametric alternative by enclosing nominal data within a minimal hypersphere in a feature space, though their kernel-based formulation can be computationally expensive.
To overcome these challenges, we propose a novel detection approach, D-RFM, which employs RFM to address both non-Gaussian clutter and additive thermal noise. Moreover, the D-RFM improves weak-target detection in low SNR.



\section{Rectified Flow-based Radar Detection}
\label{sec:method}

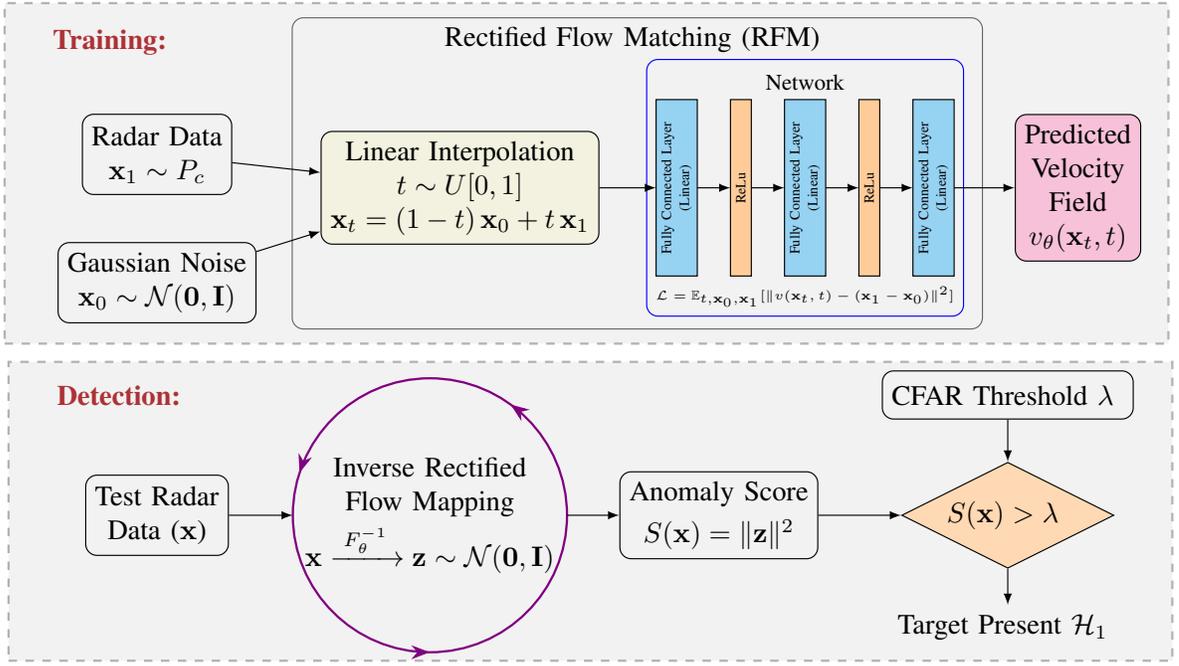
\begin{figure*}[!ht]
\centering
\resizebox{0.93\linewidth}{!}{%

\tikzstyle{input} = [coordinate]
\tikzstyle{output} = [coordinate]
\tikzstyle{block}= [draw, rectangle, rounded corners, align=center, minimum height=1cm, minimum width=1.5cm]
\tikzstyle{block1}= [draw, rectangle, align=center,minimum width=2.2cm, inner sep=2pt, outer sep=0pt]
\tikzstyle{circleblock}=[draw, circle, align=center, draw=violet, thick, minimum size=3.2cm,inner sep=0pt, outer sep=0pt]
\tikzstyle{decision}=[draw, diamond, aspect=2, align=center, inner sep=2pt]
\tikzstyle{dash_blk} = [draw, rectangle, dashed, draw=black!40, thick,align=center, minimum width=14.5cm, inner sep=3pt, outer sep=0pt] 
\pgfdeclarelayer{background}
\pgfsetlayers{background,main}

\begin{tikzpicture}

\node [input, name=input] {}; 

\node[block, right of= input, node distance=0mm] (cl) { Radar Data\\ $\mathbf{x}_1 \sim P_c$};

\node[block, below of= cl, node distance=1.6cm] (no) {Gaussian Noise\\ $\mathbf{x}_0 \sim  \mathcal{N}(\mathbf{0},\mathbf{I})$};

\node[above of= cl, node distance=1.4cm] (train) { \color{Maroon}{\textbf{Training:} \qquad \quad } };

\node[below of= train, node distance=4.4cm] (test) { \color{Maroon}{\textbf{Detection:} \qquad \ } };

\node[block, below right= -0.8cm and 1.1cm of cl, align=center,fill=olive!10] (lin) {Linear Interpolation\\ $t\sim U[0,1]$ \\ $\mathbf{x}_t=(1-t)\,\mathbf{x}_0+t\,\mathbf{x}_1$};


\node[block1, right of= lin, node distance=2.7cm, rotate=90, fill=Cerulean!40] (FL1) {\tiny{Fully Connected Layer}\\[-2.0mm] \tiny{(Linear)}};

\node[block1, right of= FL1, node distance=0.8cm, rotate=90, fill=orange!40] (RL1) {\tiny{ReLu}};

\node[block1, right of= RL1, node distance=0.8cm, rotate=90, fill=Cerulean!40] (FL2) {\tiny{Fully Connected Layer}\\[-2.0mm] \tiny{(Linear)}};

\node[block1, right of= FL2, node distance=0.8cm, rotate=90, fill=orange!40] (RL2) {\tiny{ReLu}};

\node[block1, right of=RL2, node distance=0.8cm, rotate=90, fill=Cerulean!40] (RFM) {\tiny{Fully Connected Layer}\\[-2.0mm] \tiny{(Linear)}};

\node [block, draw=blue, fit = (FL1) (RL1) (FL2) (RL2) (RFM), minimum height=3.2cm] (NN) {\footnotesize{Network}\\[2.2cm] \tiny\resizebox{\linewidth}{!}{$\mathcal{L}=\mathbb{E}_{t,\mathbf{x}_0,\mathbf{x}_1}[\|v(\mathbf{x}_t,t)-(\mathbf{x}_1-\mathbf{x}_0)\|^2]$} };

\node [block, draw=black!60, fit = (lin) (FL1) (RL1) (FL2) (RL2) (RFM), minimum height=3.88cm, minimum width=8.6cm, yshift=1.8mm ] (VFM) {Rectified Flow Matching (RFM) \vspace{4.1cm} };

\node[block, right of= RFM, node distance=1.8cm, fill=Magenta!30] (vel) {Predicted\\ Velocity\\ Field\\ $v_\theta(\mathbf{x}_t,t)$};

\begin{pgfonlayer}{background} \node [dash_blk, draw=black!30, fill=gray!10, fit = (cl) (no) (lin) (RFM) (vel) (NN) (VFM), inner sep=5pt, xshift=-1.6mm] {}; \end{pgfonlayer}

\draw [draw, ->, >=latex] (cl) --  node[above, align=center] {} (lin);
\draw [draw, ->, >=latex] (no) --  node[above, align=center] {} (lin);
\draw [draw, ->, >=latex] (lin) --  node[above, align=center] {} (FL1);
\draw [draw, ->, >=latex] (FL1) --  node[above, align=center] {} (RL1);
\draw [draw, ->, >=latex] (RL1) --  node[above, align=center] {} (FL2);
\draw [draw, ->, >=latex] (FL2) --  node[above, align=center] {} (RL2);
\draw [draw, ->, >=latex] (RL2) --  node[above, align=center] {} (RFM);
\draw [draw, ->, >=latex] (RFM) --  node[above, align=center] {} (vel);


\node[block, below of= input, node distance=4.5cm] (input1) {Test Radar\\ Data ($\mathbf{x}$)};

\node[circleblock, right of= input1, node distance=3.4cm, postaction={ decorate, decoration={ markings, mark=at position 0.15 with {\arrow{Stealth}}, mark=at position 0.45 with {\arrow{Stealth}}, mark=at position 0.75 with {\arrow{Stealth}} } }] (inverse) { Inverse Rectified \\ Flow Mapping \\[1mm]   $\mathbf{x} \xrightarrow{\;F_\theta^{-1}\;} \mathbf{z} \sim \mathcal{N}(\mathbf{0},\mathbf{I})$ };

\node[block, right of= inverse, node distance=3.6cm] (score) {Anomaly Score \\[1mm]    $S(\mathbf{x}) = \|\mathbf{z}\|^2$ };

\node[decision, right of= score, node distance=3.6cm,fill=orange!30] (decision) { $S(\mathbf{x}) > \lambda$ };

\node[block, above of =decision, node distance=1.5cm, align=center, minimum height=0.6cm] (thr) {CFAR Threshold $\lambda$ };

\node[below of= decision, node distance=1.4cm, align=center] (tar) {Target Present $\mathcal{H}_1$ };

\begin{pgfonlayer}{background} \node [dash_blk, draw=black!30, fill=gray!10, fit = (input1) (test) (thr) (inverse) (score) (decision) (tar) ] {}; \end{pgfonlayer}

\draw [draw, ->, >=latex] (input1) --  node[above, align=center] {} (inverse);
\draw [draw, ->, >=latex] (inverse) --  node[above, align=center] {} (score);
\draw [draw, ->, >=latex] (score) --  node[above, align=center] {} (decision);
\draw [draw, ->, >=latex] (thr) --  node[above, align=center] {} (decision);
\draw [draw, ->, >=latex] (decision) --  node[above, align=center] {} (tar);

\end{tikzpicture}
}
\caption{Overall framework for proposed rectified flow matching-based radar target detection. }
\label{fig:method}
\end{figure*}

This section presents the proposed D-RFM framework, as shown in Fig.~\ref{fig:method}. The framework consists of \textit{Training} and \textit{Detection} stages. The \textit{Training} stage involves learning a mapping from a standard Gaussian distribution to radar clutter-plus-noise samples using the RFM network. In the \textit{Detection} stage, a new or test radar observation is mapped back into the latent Gaussian space via the learned mapping, and an anomaly score is computed to identify the presence of a target. Further details of the two stages are provided below. 

\subsection{Training}
\label{sec:train}

The goal of the training stage is to learn a mapping that transforms a standard Gaussian distribution into radar observations under $\mathcal{H}_0$ (clutter-plus-noise), enabling the detection of anomalies corresponding to targets. We employed RFM, which provides a simple and effective approach to learn mapping by modeling the velocity field along linear interpolations between Gaussian noise and radar observations. 

Let $P_c$ denote the distribution of radar observations under hypothesis $\mathcal{H}_0$, and $\mathbf{y} \sim P_c$ denote a radar sample. 
 Since $\mathbf{y}$ is complex-valued, each observation is represented as a real-valued vector $\mathbf{x_1} \in \mathbb{R}^D$ by concatenating the real and imaginary parts $\mathbf{x}_1 = \big[ \Re\{\mathbf{y}\}, \; \Im\{\mathbf{y}\} \big]^T \in \mathbb{R}^D$. 
  Here, $D = 2 N$ accounts for both the real and imaginary components, and $N$ is the number of radar pulses. Similarly, let $\mathcal{N}(\mathbf{0}, \mathbf{I})$ denote a standard Gaussian distribution, and $\mathbf{x}_0 \sim \mathcal{N}(\mathbf{0}, \mathbf{I})$ be a latent noise vector with the same dimension as $\mathbf{x}_1$. For each training sample, the linear interpolation between the observed radar vector $\mathbf{x}_1$ and latent Gaussian vector $\mathbf{x}_0$ is defined as:
\begin{equation}
\mathbf{x}_t = (1-t)\,\mathbf{x}_0 + t\,\mathbf{x}_1, \quad t\in [0;1]\, ,
\label{eq:lnpath}
\end{equation}
where $t$ is a temporal interpolation variable sampled uniformly from $[0;1]$. Here, $t=0$ corresponds to the Gaussian latent vector, and $t=1$ corresponds to the radar observation.


Given $\mathbf{x}_0 \sim \mathcal{N}(\mathbf{0}, \mathbf{I})$ and $\mathbf{x}_1 \sim P_c$, the rectified flow network aims to learn a time-dependent velocity field $\mathbf{v}_\theta: \mathbb{R}^D \times [0;1] \rightarrow \mathbb{R}^D$, that approximates the constant direction along the linear interpolation: $\mathbf{v}_\theta(\mathbf{x}_t, t) \approx \mathbf{x}_1 - \mathbf{x}_0$ and this velocity field $\mathbf{v}_\theta$, referred to as rectified flow. The network is implemented using three fully connected layers with ReLU activations and is trained by minimizing the loss between the predicted velocity and the true interpolation direction as:
\begin{equation}
    \mathcal{L}(\theta) = \mathbb{E}_{t, \mathbf{x}_0 \sim \mathcal{N}(\mathbf{0}, \mathbf{I}),\, \mathbf{x}_1 \sim P_c} \Big[ \|  \mathbf{v}_\theta(\mathbf{x}_t, t) - (\mathbf{x}_1 - \mathbf{x}_0) \|^2 \Big]\, 
\label{eq:rfmloss}
\end{equation}

Training uses the Adam optimizer with a learning rate of $1 \times 10^{-3}$, a batch size of $128$, and $170$ epochs. These values were chosen based on empirical experiments.
Minimizing this loss allows the network to learn a rectified flow $\mathbf{v}_\theta$ that maps $\mathbf{x}_0$ to $\mathbf{x}_1$.
latent Gaussian samples to radar clutter-plus-noise samples. The learned flow is later reversed during detection to identify the presence of the target.

\subsection{Detection}

Detection is performed by mapping a test radar sample back into the latent Gaussian space and evaluating its deviation from the learned distribution. 

Let $\mathbf{y}_\text{test}$ be a new radar observation and $\mathbf{x} \in \mathbb{R}^D$ be its real-valued representation obtained by concatenating the real and imaginary parts. Then the inverse mapping is performed using the learned rectified flow (i.e., velocity field):
\begin{equation}
\mathbf{z} = F_\theta^{-1}(\mathbf{x}) 
\sim \mathcal{N}(\mathbf{0}, \mathbf{I})\,,
\end{equation}
\noindent
where $\mathbf{z}$ is the latent vector obtained after inverse mapping. $F_\theta^{-1}$ represents the inverse mapping from $t=1$ to $t=0$ induced by the learned velocity field $\mathbf{v}_\theta$ and defined as $\mathbf{z} = \mathbf{x} - \displaystyle\int_0^1 \mathbf{v}(\mathbf{x}_t, t) \, dt$. 

Under hypothesis $\mathcal{H}_0$, the latent vector $\mathbf{z}$ is expected to follow a standard Gaussian distribution. If a target is present ($\mathcal{H}_1$), $\mathbf{z}$ deviates from the Gaussian distribution due to the contribution of the target signal. To quantify this deviation, an anomaly score is defined as the squared $\ell_2$-norm of the latent vector: $S(\mathbf{x}) = \|\mathbf{z}\|^2$. Detection is then performed by comparing $S(\mathbf{x})$ 
with a CFAR threshold $\lambda$, determined according to a desired false alarm probability:
\begin{equation}
\hat{\mathcal{H}} =
\begin{cases}
\mathcal{H}_0, & S(\mathbf{x}) \leq \lambda, \\
\mathcal{H}_1, & S(\mathbf{x}) > \lambda.
\end{cases}
\end{equation}

The proposed D-RFM framework is robust to both Gaussian and non-Gaussian clutter, as it learns the true underlying distribution of clutter-plus-noise. Additionally, the use of linear interpolation and network-based velocity prediction ensures computational efficiency.


\section{Results and Discussion}
\label{sec:results}

This section presents a performance analysis of the proposed D-RFM and compares it with classical detectors, including MF, NMF, AMF-SCM, ANMF-FP, ANMF-SCM, and SVDD. The performance of each detector is evaluated under two scenarios: correlated Gaussian noise with additive white Gaussian noise (cGN + AWGN), and correlated compound Gaussian noise with additive white Gaussian noise (cCGN + AWGN). The quantitative measure used to assess the detection performance is probability of detection ($P_d$) as a function of SNR, with $P_{fa}$ = $10^{-2}$. Here, $P_{fa}$ represents the false-alarm probability. 



\subsection{Experimental Setup:} 
The target echo is modeled as $\alpha = \sqrt{\frac{\mathrm{SNR}}{N}}\, e^{2j\pi \phi}$, where $\phi \in [0;1]$. The steering vector is denoted by $\mathbf{p} = \left(1,\, e^{2j\pi d/N},\, \ldots,\, e^{2j\pi d(N-1)/N} \right)^{T}$. Here, $N$ and $d$ represent the number of Doppler bins and the normalized Doppler bin index of the target. In experiments, we set $N = 16$. 
The clutter covariance matrix is modeled as $\boldsymbol{\Sigma}_c = \mathcal{T}(\rho)$, with a correlation coefficient $\rho = 0.5$. 
The texture components $\delta$ is drawn from a Gamma distribution $\Gamma(\mu, 1/\mu)$ with $\mu = 1$. 
For adaptive detectors, the covariance matrix is determined via SCM or the Tyler estimator, with $K = 2N$ independent secondary data samples. 
For training the RFM detector, radar data comprising clutter-plus-noise Doppler profiles is used for each noise scenario. 
The dataset used is split into $10000$ training samples, $10000$ validation samples (used to fix the CFAR threshold $\lambda$), and $5000$ test samples.
The other details of the hyperparameters are provided in Section \ref{sec:train}. 

\pgfplotsset{
    tiny,
    legend style={
         at={(0.07,0.065)},
        anchor=south west,
    },
    every axis plot/.append style={line width=1.0pt},
   }%

\begin{figure}[!hbt]
\centering
\begin{tikzpicture}[baseline]
\begin{axis}[
    width=1\linewidth, 
    height=65mm,
    xlabel={\small{SNR (dB)}},
    xlabel style={yshift=5pt}, 
    ylabel={\normalsize{$p_d$}},
    ylabel style={yshift=-3pt}, 
    xmin=-20, xmax=20,
    ymin=-0.1, ymax=1.05,
    xtick={-20,-15,-10,-5,0,5,10,15,20},
    ytick={0,0.2,0.4,0.6,0.8,1},
    tick label style={font=\small},  
    legend pos=north west,
    legend columns=1, 
    legend style={draw=none, font=\small, fill=none},
    ymajorgrids=true,
    xmajorgrids=true,
    grid style=dashed,
]

\addplot[  color=blue,
    ]
    coordinates {(-20,0.0136) (-19,0.0144) (-18,0.0136) (-17,0.0122) (-16,0.0126) (-15,0.0134) (-14,0.0146) (-13,0.0148) (-12,0.0156) (-11,0.0176) (-10,0.021) (-9,0.024) (-8,0.027) (-7,0.028) (-6,0.029) (-5,0.031) (-4,0.032) (-3,0.037) (-2,0.041) (-1,0.049) (0,0.052) (1,0.066) (2,0.084) (3,0.108) (4,0.138) (5,0.168) (6,0.215) (7,0.285) (8,0.375) (9,0.489) (10,0.629) (11,0.786) (12,0.911) (13,0.969) (14,0.998) (15,1.0) (16,1.0) (17,1.0) (18,1.0) (19,1.0)       };
    \addlegendentry{D-RFM}
\addplot[    color=violet, 
    ]
    coordinates {(-20,0.0124) (-19,0.0124) (-18,0.0144) (-17,0.0132) (-16,0.0124) (-15,0.012) (-14,0.011) (-13,0.012) (-12,0.0126) (-11,0.0108) (-10,0.0122) (-9,0.012) (-8,0.0122) (-7,0.0116) (-6,0.015) (-5,0.0122) (-4,0.0152) (-3,0.014) (-2,0.0152) (-1,0.0174) (0,0.018) (1,0.0226) (2,0.0264) (3,0.036) (4,0.0504) (5,0.0648) (6,0.1086) (7,0.1568) (8,0.247) (9,0.3684) (10,0.5398) (11,0.718) (12,0.868) (13,0.956) (14,0.9916) (15,0.9998) (16,0.9998) (17,1.0) (18,1.0) (19,1.0)
    };
    \addlegendentry{SVDD}
\addplot[    color=cyan
    ]
    coordinates {  (-20, 0.0148) (-19, 0.0152) (-18, 0.0152) (-17, 0.0154) (-16, 0.0158) (-15, 0.0158) (-14, 0.0158) (-13, 0.017) (-12, 0.0174) (-11, 0.0188) (-10, 0.0202) (-9, 0.0206) (-8, 0.0236) (-7, 0.0264) (-6, 0.0302) (-5, 0.035) (-4, 0.04) (-3, 0.0484) (-2, 0.062) (-1, 0.0752) (0, 0.0994) (1, 0.1362) (2, 0.1804) (3, 0.235) (4, 0.3062) (5, 0.4032) (6, 0.5186) (7, 0.6502) (8, 0.7782) (9, 0.8824) (10, 0.9556) (11, 0.9886) (12, 0.9978) (13, 1) (14, 1) (15, 1) (16, 1) (17, 1) (18, 1) (19, 1)    };


    \addlegendentry{MF - oracle}
\addplot[    color=Magenta
    ]
    coordinates {  (-20, 0.0096) (-19, 0.0096) (-18, 0.01) (-17, 0.01) (-16, 0.0102) (-15, 0.0102) (-14, 0.0106) (-13, 0.0112) (-12, 0.0116) (-11, 0.0128) (-10, 0.0142) (-9, 0.0148) (-8, 0.0162) (-7, 0.0182) (-6, 0.0206) (-5, 0.0244) (-4, 0.0302) (-3, 0.0344) (-2, 0.042) (-1, 0.053) (0, 0.0716) (1, 0.0962) (2, 0.1248) (3, 0.1736) (4, 0.2296) (5, 0.3086) (6, 0.4068) (7, 0.5292) (8, 0.6616) (9, 0.7894) (10, 0.8976) (11, 0.9642) (12, 0.9898) (13, 0.9996) (14, 1) (15, 1) (16, 1) (17, 1) (18, 1) (19, 1)    };
    \addlegendentry{NMF - oracle}
\addplot[    color=black, 
    ]
    coordinates {  (-20, 0.0082) (-19, 0.0082) (-18, 0.0084) (-17, 0.0084) (-16, 0.0084) (-15, 0.0082) (-14, 0.0086) (-13, 0.0096) (-12, 0.0102) (-11, 0.0102) (-10, 0.0104) (-9, 0.0104) (-8, 0.011) (-7, 0.0114) (-6, 0.0116) (-5, 0.0122) (-4, 0.0134) (-3, 0.016) (-2, 0.0194) (-1, 0.0236) (0, 0.0302) (1, 0.0368) (2, 0.0494) (3, 0.0634) (4, 0.086) (5, 0.1202) (6, 0.16) (7, 0.2174) (8, 0.3018) (9, 0.4132) (10, 0.5468) (11, 0.6918) (12, 0.8136) (13, 0.9088) (14, 0.9676) (15, 0.9936) (16, 0.9996) (17, 1) (18, 1) (19, 1)    };
    \addlegendentry{AMF-SCM}
\addplot[    color=red, 
    ]
    coordinates { (-20, 0.0086) (-19, 0.0086) (-18, 0.0088) (-17, 0.0084) (-16, 0.0086) (-15, 0.0088) (-14, 0.009) (-13, 0.0096) (-12, 0.0094) (-11, 0.0096) (-10, 0.01) (-9, 0.0106) (-8, 0.0106) (-7, 0.012) (-6, 0.0138) (-5, 0.0158) (-4, 0.0188) (-3, 0.0224) (-2, 0.0256) (-1, 0.0294) (0, 0.0378) (1, 0.0476) (2, 0.0618) (3, 0.0794) (4, 0.1068) (5, 0.14) (6, 0.1806) (7, 0.2432) (8, 0.321) (9, 0.4202) (10, 0.5508) (11, 0.6722) (12, 0.7822) (13, 0.8768) (14, 0.9388) (15, 0.9756) (16, 0.9936) (17, 0.999) (18, 0.9998) (19, 1)    };
    \addlegendentry{ANMF-FP}
\addplot[    color=OliveGreen, 
    dashed
    ]
    coordinates { (-20, 0.009) (-19, 0.0088) (-18, 0.0088) (-17, 0.0086) (-16, 0.0084) (-15, 0.0086) (-14, 0.0084) (-13, 0.0086) (-12, 0.0094) (-11, 0.0096) (-10, 0.01) (-9, 0.0102) (-8, 0.0108) (-7, 0.0126) (-6, 0.0138) (-5, 0.0158) (-4, 0.0174) (-3, 0.0202) (-2, 0.024) (-1, 0.031) (0, 0.038) (1, 0.0478) (2, 0.0612) (3, 0.0792) (4, 0.1048) (5, 0.1378) (6, 0.1822) (7, 0.2424) (8, 0.325) (9, 0.426) (10, 0.552) (11, 0.68) (12, 0.7868) (13, 0.8806) (14, 0.9444) (15, 0.9772) (16, 0.9938) (17, 0.999) (18, 0.9998) (19, 1)    };
    \addlegendentry{ANMF-SCM}
\end{axis}
\end{tikzpicture}

\centerline{ (a) cGN + AWGN  }
\vspace{2mm}

\begin{tikzpicture}[baseline]
\begin{axis}[
    width=1\linewidth, 
    height=64mm,
    xlabel={\small{SNR (dB)}},
    xlabel style={yshift=5pt}, 
    ylabel={\normalsize{$p_d$}},
    ylabel style={yshift=-3pt}, 
    xmin=-20, xmax=20,
    ymin=-0.1, ymax=1.05,
    xtick={-20,-15,-10,-5,0,5,10,15,20},
    ytick={0,0.2,0.4,0.6,0.8,1},
    tick label style={font=\small},
    legend pos=north west,
    legend columns=1, 
    legend style={draw=none, font=\small, fill=none},
    ymajorgrids=true,
    xmajorgrids=true,
    grid style=dashed,
]

\addplot[
    color=blue,
    ]
    coordinates { (-20,0.0120) (-19,0.0126) (-18,0.0124) (-17,0.0152) (-16,0.0136) (-15,0.0122) (-14,0.0124) (-13,0.0138) (-12,0.0118) (-11,0.0116) (-10,0.0170) (-9,0.0166) (-8,0.0186) (-7,0.0164) (-6,0.0178) (-5,0.0174) (-4,0.0190) (-3,0.0220) (-2,0.0254) (-1,0.0330) (0,0.0320) (1,0.0396) (2,0.0494) (3,0.0694) (4,0.0946) (5,0.1140) (6,0.1688) (7,0.2270) (8,0.3156) (9,0.4256) (10,0.5844) (11,0.7442) (12,0.8798) (13,0.9606) (14,0.9902) (15,0.9986) (16,1.0000) (17,1.0000) (18,1.0000) (19,1.0000)
    };
    \addlegendentry{D-RFM}
\addplot[
    color=violet, 
    ]
    coordinates { (-20,0.0004) (-19,0.0010) (-18,0.0010) (-17,0.0010) (-16,0.0010) (-15,0.0008) (-14,0.0004) (-13,0.0004) (-12,0.0002) (-11,0.0012) (-10,0.0004) (-9,0.0002) (-8,0.0018) (-7,0.0008) (-6,0.0010) (-5,0.0008) (-4,0.0002) (-3,0.0008) (-2,0.0008) (-1,0.0008) (0,0.0010) (1,0.0024) (2,0.0018) (3,0.0018) (4,0.0022) (5,0.0042) (6,0.0084) (7,0.0086) (8,0.0242) (9,0.0478) (10,0.1000) (11,0.2124) (12,0.4104) (13,0.6520) (14,0.8700) (15,0.9752) (16,0.9966) (17,1.0000) (18,1.0000) (19,1.0000)
    };
    \addlegendentry{SVDD}

\addplot[
    color=black, 
    ]
    coordinates { (-20,0.0138) (-19,0.014) (-18,0.014) (-17,0.0142) (-16,0.0142) (-15,0.0142) (-14,0.014) (-13,0.0144) (-12,0.0146) (-11,0.0146) (-10,0.0148) (-9,0.0154) (-8,0.0158) (-7,0.0158) (-6,0.0168) (-5,0.0184) (-4,0.0198) (-3,0.0214) (-2,0.0236) (-1,0.0254) (0,0.0292) (1,0.0332) (2,0.039) (3,0.0464) (4,0.0612) (5,0.0796) (6,0.1066) (7,0.1428) (8,0.1978) (9,0.259) (10,0.3364) (11,0.4322) (12,0.5342) (13,0.6374) (14,0.7334) (15,0.8196) (16,0.8768) (17,0.9272) (18,0.965) (19,0.9876)
    };

    \addlegendentry{AMF-SCM}
\addplot[
    color=red, 
    ]
    coordinates { (-20,0.0084) (-19,0.0084) (-18,0.0084) (-17,0.009) (-16,0.009) (-15,0.0092) (-14,0.0092) (-13,0.0096) (-12,0.01) (-11,0.011) (-10,0.0114) (-9,0.012) (-8,0.013) (-7,0.014) (-6,0.0146) (-5,0.0172) (-4,0.0186) (-3,0.0216) (-2,0.0272) (-1,0.0322) (0,0.0408) (1,0.0518) (2,0.0686) (3,0.0908) (4,0.1194) (5,0.16) (6,0.2128) (7,0.2804) (8,0.3588) (9,0.4616) (10,0.5704) (11,0.67) (12,0.7596) (13,0.8404) (14,0.9002) (15,0.9444) (16,0.9714) (17,0.9886) (18,0.996) (19,0.998)
    };
    \addlegendentry{ANMF-FP}
\end{axis}
\end{tikzpicture}
\centerline{ (b) cCGN + AWGN }
\caption{Performance comparison in terms of $p_d$, under different noise for Doppler bin $d = 0$ ($P_{fa}$ = $10^{-2}$). }
\label{fig:plots}
\end{figure}
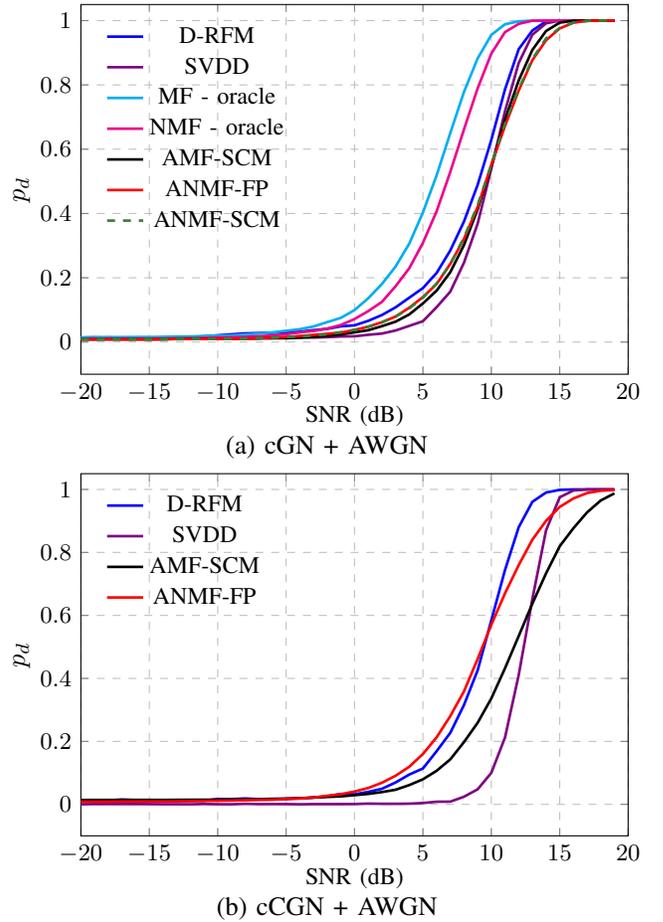


\subsection{SNR vs \texorpdfstring{$P_d$}{Pd} analysis for zero Doppler bin}

Fig.~\ref{fig:plots} compares the detection performance of D-RFM with AMF-SCM, ANMF-FP, ANMF-SCM, and SVDD for the zero-Doppler bin, with MF and NMF used as reference detectors (known as oracles in Fig. \ref{fig:plots}(a)). As shown in the figure, ANMF-FP, ANMF-SCM, and AMF-SCM exhibit comparable detection performance under the cGN+AWGN scenario. MF and NMF achieve the best detection performance among all evaluated methods. SVDD exhibits lower detection performance at low SNR levels but achieves improved performance at higher SNR levels, and surpasses ANMF-FP, ANMF-SCM, and AMF-SCM. D-RFM consistently outperforms ANMF-FP, ANMF-SCM, AMF-SCM, and SVDD across the entire SNR range. Furthermore, D-RFM achieves competitive performance relative to the MF and NMF reference detectors at both low and high SNR levels, while slightly underperforming in the mid-SNR range under the cGN+AWGN scenario.

In the cCGN+AWGN scenario, since the distribution of the interferences cannot be characterized, the oracle cannot be designed. In this context, AMF-SCM exhibits the lowest detection performance. D-RFM outperforms SVDD and shows performance comparable to ANMF-FP at low SNR levels. As the SNR increases, D-RFM surpasses ANMF-FP and remains superior to the other evaluated methods, indicating that although target detection is more challenging at low SNRs, the proposed detector adapts effectively and achieves competitive performance at higher SNR levels.

\subsection{SNR vs \texorpdfstring{$P_d$}{Pd} analysis for all Doppler bins}

We further assessed the detection performance of the methods across all Doppler bins, as shown in Fig.~\ref{fig:intersection}. The figure shows that D-RFM outperforms both SVDD and AMF-SCM across all bins in the cGN+AWGN scenario. This demonstrates the efficacy of D-RFM in handling noise. 

For the cCGN+AWGN scenario, ANMF-FP outperforms both SVDD and D-RFM at low SNR levels. However, D-RFM achieves performance comparable to ANMF-FP across mid-to-high SNR levels. At mid SNR ($10 - 12$dB), D-RFM’s $P_d$ values are slightly lower than ANMF-FP in some Doppler bins. As SNR increases beyond $12 - 15$ dB, D-RFM reaches $P_d \approx 1$ across almost all Doppler bins, often slightly earlier than ANMF-FP in several bins. Overall, D-RFM exhibits slightly inferior performance at lower SNRs, while surpassing ANMF-FP at the highest SNR values. These results demonstrate the robustness of D-RFM across all Doppler bins and SNR levels.

\tikzstyle{block} = [draw, rectangle, minimum height=3em, minimum width=6em]   
\tikzstyle{block1} = [draw=none, thick, minimum height=3em, minimum width=6em]
\tikzstyle{input} = [coordinate]
\tikzstyle{output} = [coordinate]
\tikzstyle{dotted_block}=[draw, rectangle, line width=1pt, dash pattern=on 1.5pt off 3pt on 6.5pt off 3pt, rounded corners]

\begin{figure}[!t]
\centering
\resizebox{\linewidth}{!}{%
\begin{tikzpicture}
 
\node [input, name=input] {};


\node [right of=input,  node distance=-3.8cm, align=center] (a1) {\includegraphics[width=0.5\linewidth, height=4.9cm]{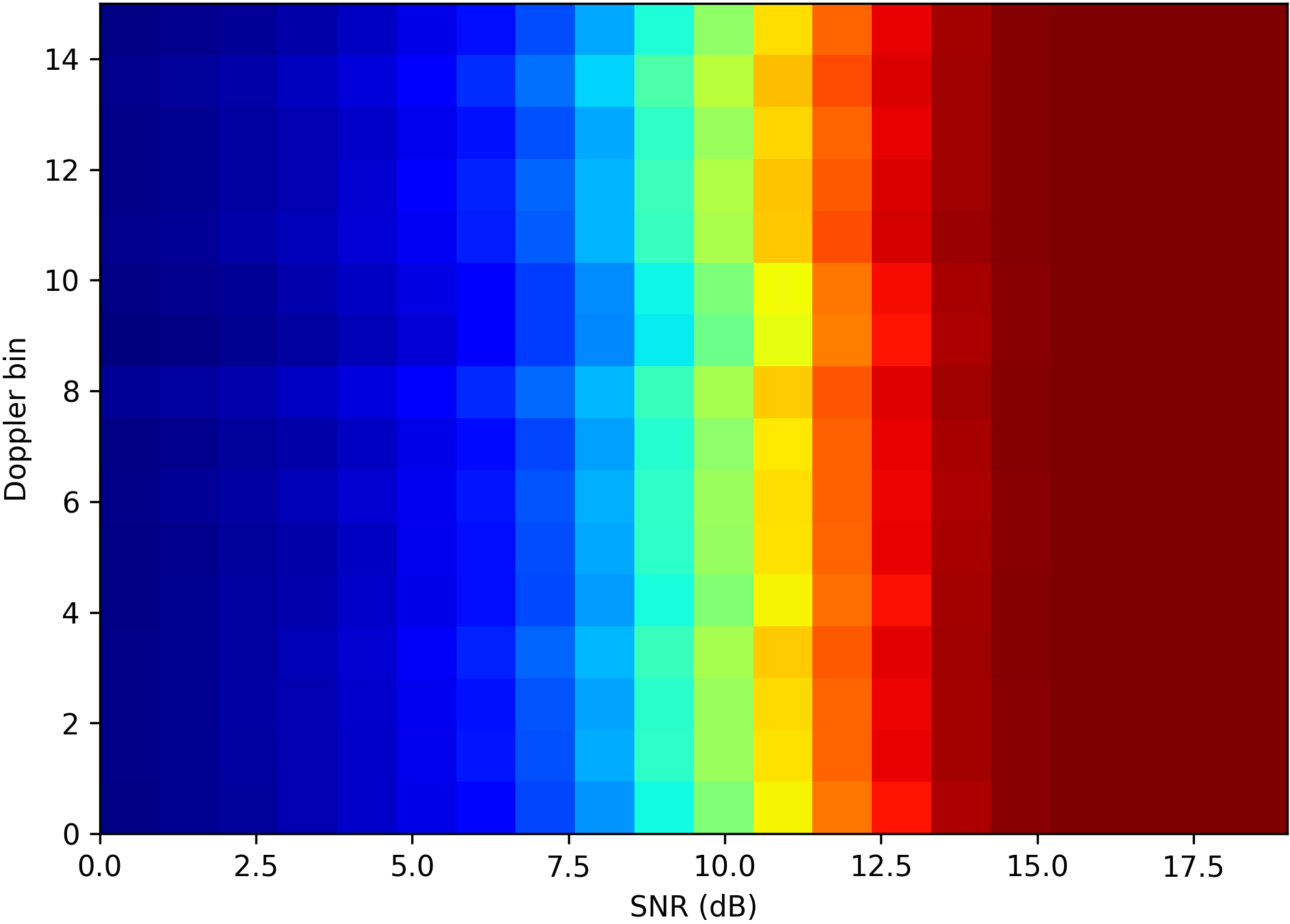} };
\node[ below=-0.3cm of a1, fill=white, inner sep=2pt ] {\footnotesize{SNR (dB)}};
\node[ rotate=90, left=-0.1cm of a1, fill=white, inner sep=2pt, xshift=0.9cm] {\footnotesize{Doppler bin}};

\node [right of=a1,  node distance=4.8cm, align=center] (a2) { \includegraphics[width=0.5\linewidth, height=4.9cm]{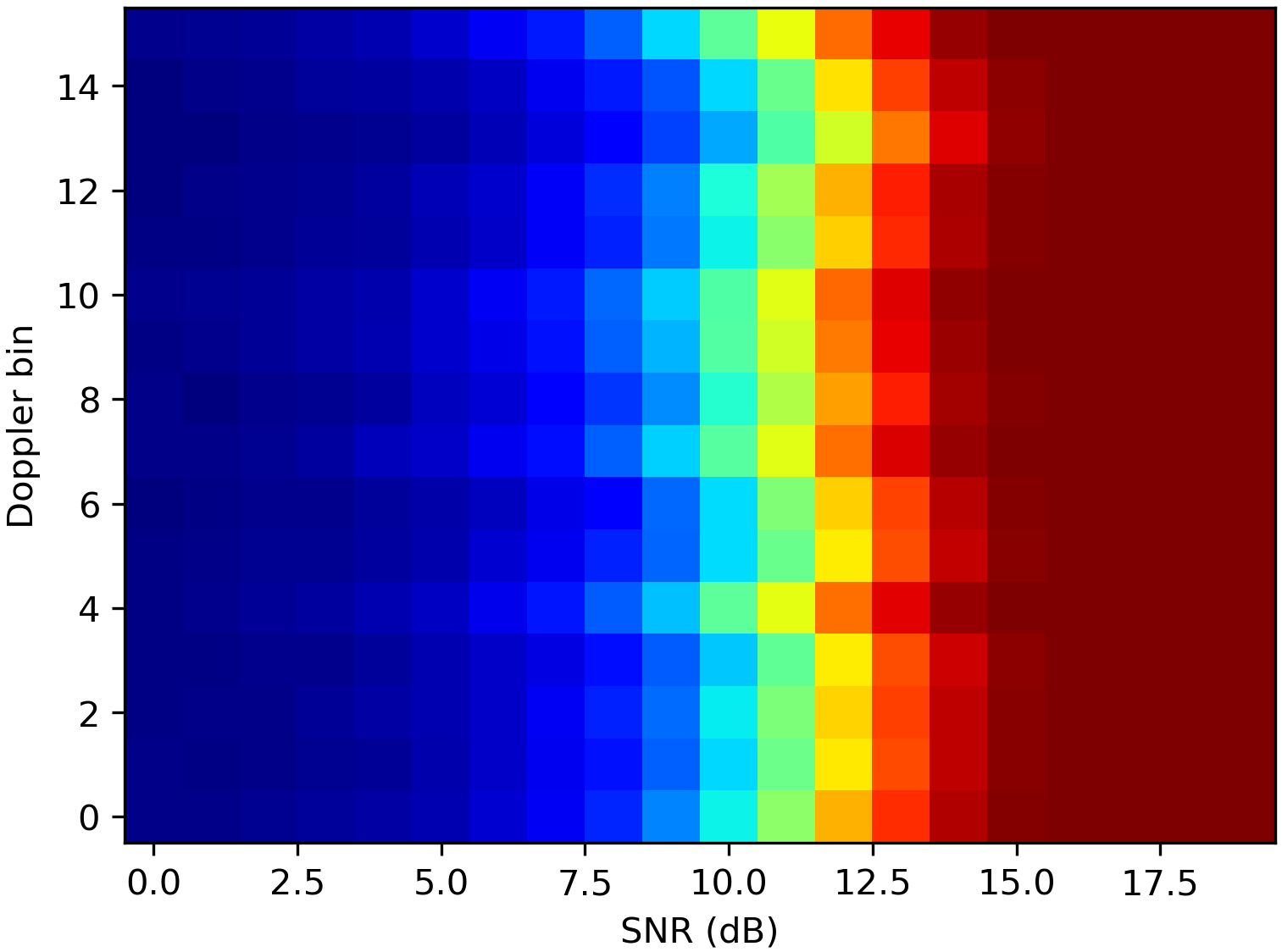}}; 
\node[ below=-0.3cm of a2, fill=white, inner sep=2pt ] {\footnotesize{SNR (dB)}};
\node[ rotate=90, left=-0.1cm of a2, fill=white, inner sep=2pt, xshift=0.9cm] {\footnotesize{Doppler bin}};

\node [right of=a2,  node distance=5.1cm, align=center] (a3) {\includegraphics[width=0.56\linewidth, height=4.9cm]{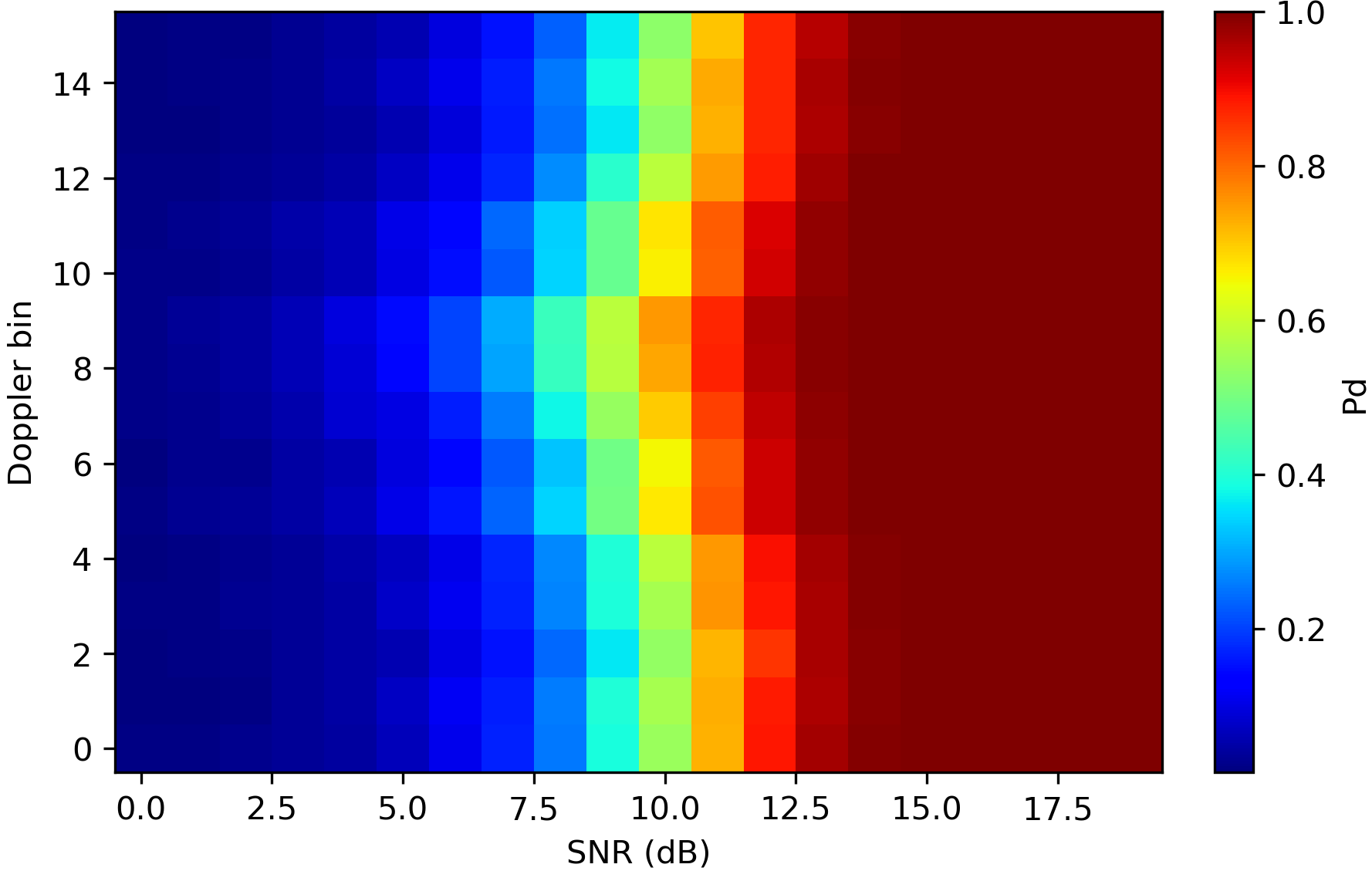} };
\node[ below=-0.32cm of a3, fill=white, inner sep=2pt ] {\footnotesize{SNR (dB)}};
\node[ rotate=90, left=-0.1cm of a3, fill=white, inner sep=2pt, xshift=0.9cm] {\footnotesize{Doppler bin}};

\node [below of=a2,  node distance=2.9cm, align=center] (alabel){ \hspace{-0.4cm} AMF-SCM \hspace{3.3cm}  SVDD \hspace{3.5cm}  D-RFM};

 
 \node [below of=a1,  node distance=5.6cm, align=center] (b1) {\includegraphics[width=0.5\linewidth, height=4.9cm]{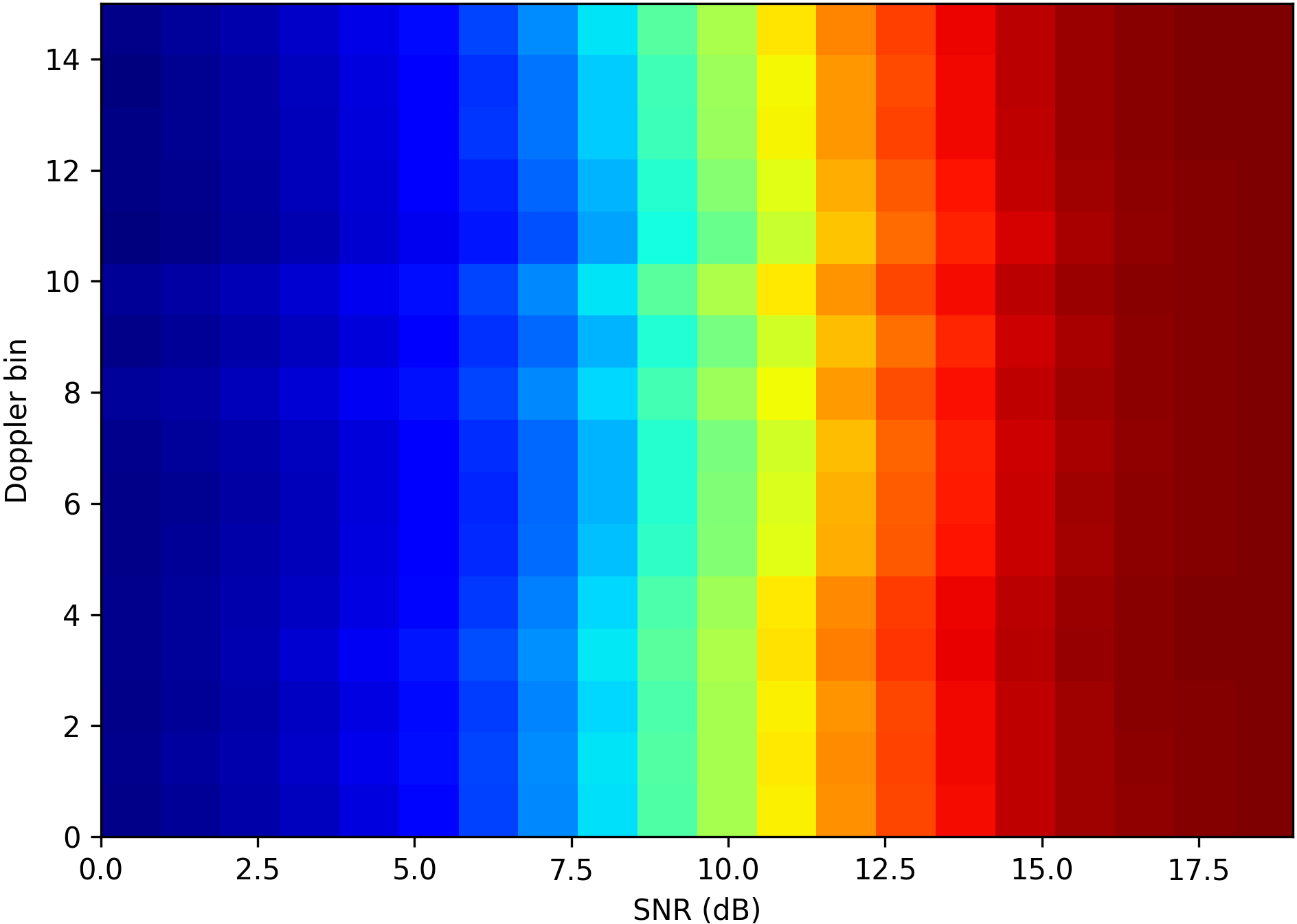} };
\node[ below=-0.3cm of b1, fill=white, inner sep=2pt ] {\footnotesize{SNR (dB)}};
\node[ rotate=90, left=-0.1cm of b1, fill=white, inner sep=2pt, xshift=0.9cm] {\footnotesize{Doppler bin}};

\node [right of=b1,  node distance=4.8cm, align=center] (b2) { \includegraphics[width=0.5\linewidth, height=4.9cm]{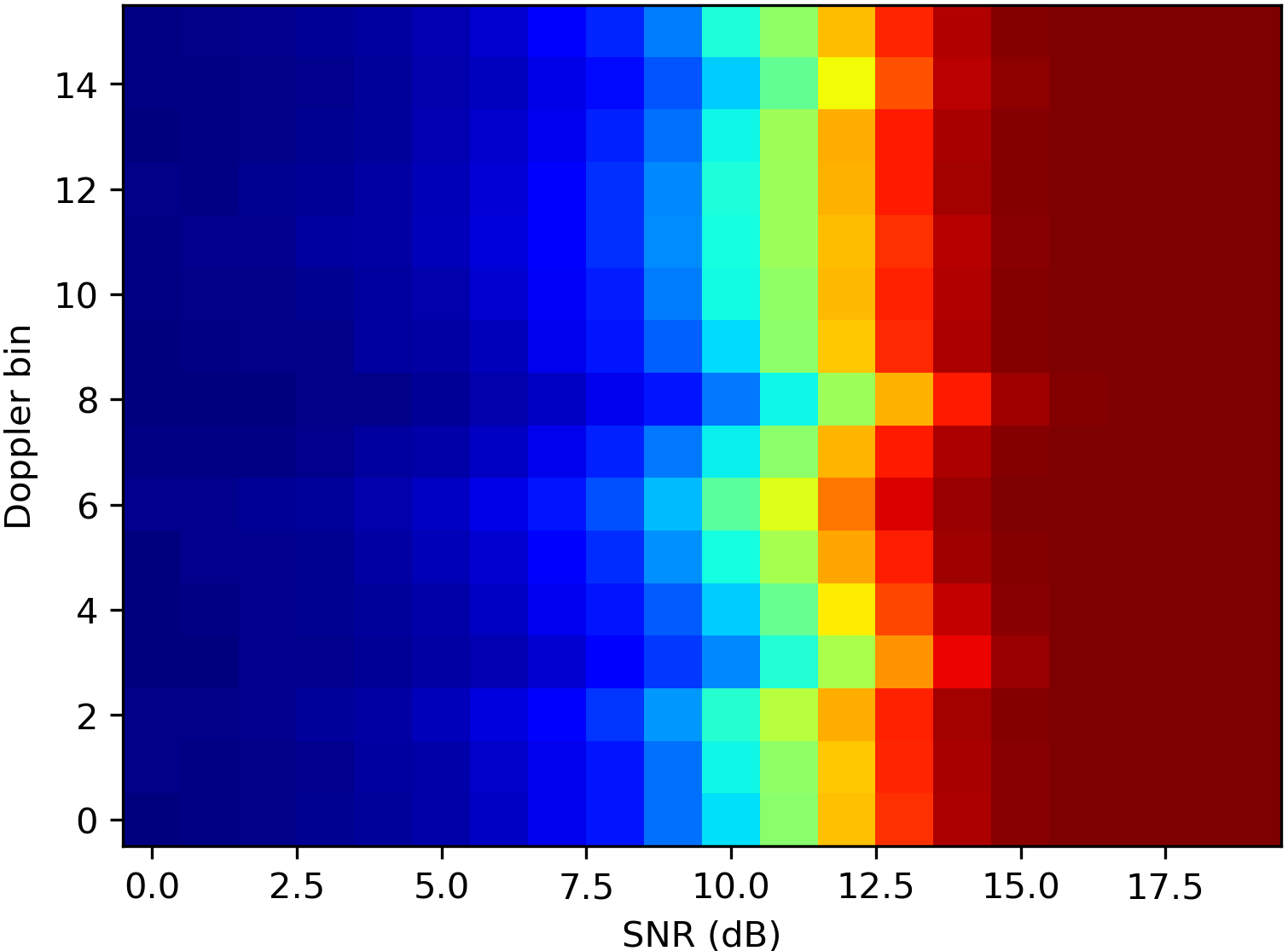} }; 
\node[ below=-0.3cm of b2, fill=white, inner sep=2pt ] {\footnotesize{SNR (dB)}};
\node[ rotate=90, left=-0.1cm of b2, fill=white, inner sep=2pt, xshift=0.9cm] {\footnotesize{Doppler bin}};

\node [right of=b2,  node distance=5.1cm, align=center] (b3) {\includegraphics[width=0.56\linewidth, height=4.9cm]{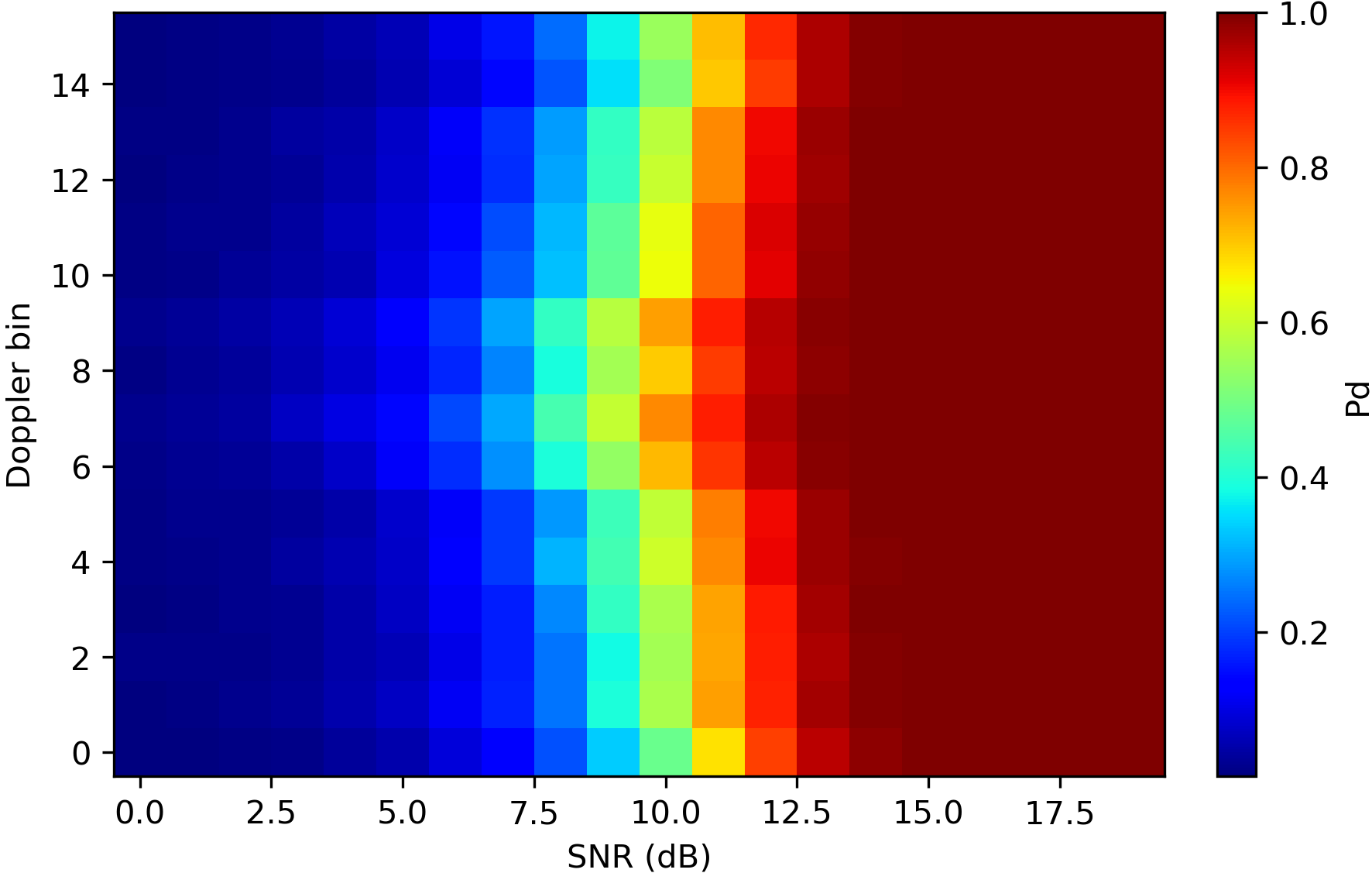}};
\node[ below=-0.32cm of b3, fill=white, inner sep=2pt ] {\footnotesize{SNR (dB)}};
\node[ rotate=90, left=-0.1cm of b3, fill=white, inner sep=2pt, xshift=0.9cm] {\footnotesize{Doppler bin}};

\node [below of=b2,  node distance=2.9cm, align=center] (blabel){ \hspace{-0.4cm} ANMF-FP \hspace{3.4cm}  SVDD \hspace{3.4cm}  D-RFM};
\end{tikzpicture}
}   
\caption{Comparison of $P_d$-SNR-Doppler bin map obtained using AMF-SCM, SVDD, ANMF-FP, and D-RFM for different cases. (Top row): cGN + AWGN (Bottom row): cCGN + AWGN. }
\label{fig:intersection}
\end{figure}
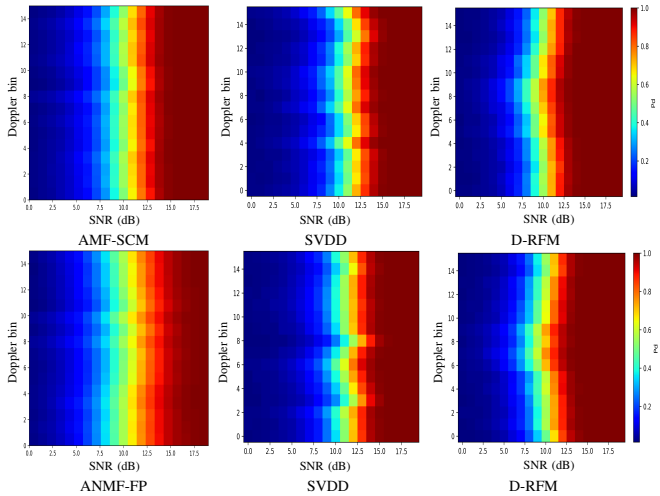

\subsection{Computational complexity analysis}
We also evaluate the computational complexity of the D-RFM method against existing methods. Table \ref{tab:time_comparison} compares the average per-sample detection time of the different detectors. The time is calculated as $\bar{t} = \frac{1}{\mathcal{S}} \sum_{i=1}^{\mathcal{S}} \hat{t}_i$, where $\hat{t}_i = \frac{T_i}{\mathcal{N}}$, $\mathcal{S}$ is the number of SNR values considered ($0$-$20$dB), $T_i$ is the total detection time over all test samples at the $i$-th SNR for a given Doppler bin, and $\mathcal{N}$ is the number of test samples per SNR. In our experiments, $\mathcal{N} = 1000$ for a single Doppler bin. As shown in the table, D-RFM achieves the lowest per-sample detection time among all evaluated methods, demonstrating its computational efficiency while providing robust performance under non-Gaussian clutter and improved weak-target detection.

\vspace{-0.2cm}

\begin{table}[htbp]
\centering
\caption{Performance comparison in terms of average per-sample detection time $\bar{t}$ ($\mathrm{ms}$).}
\label{tab:time_comparison}
\resizebox{\columnwidth}{!}{%
\begin{tabular}{lcccccc}
\hline
\textbf{Method}       & MF & NMF &  AMF-SCM & ANMF-FP& SVDD & D-RFM \\
 \hline
\textbf{Time} ($\bar{t}$) & $0.0516$ & $0.1376$ & $0.0970$ & $1.9255$ & $0.4042$ &  $\mathbf{0.0209}$ \\
\hline
\end{tabular}
}
\end{table}



\vspace{-0.4cm}
\section{Conclusion}

This paper introduced D-RFM, an RFM-based framework for radar target detection in challenging clutter environments. The proposed approach learns a deterministic flow that maps a standard Gaussian distribution to clutter-plus-noise observations, enabling target detection as deviations from the learned distribution. By combining linear interpolation with a neural network–parameterized velocity field, D-RFM provides robustness to non-Gaussian clutter while maintaining low computational complexity. Experimental results show that the proposed detector outperforms conventional methods and SVDD across a wide range of SNRs while preserving fast per-sample inference time, highlighting its potential for weak-target detection in heterogeneous radar scenarios.
However, the method assumes that target presence induces increased energy in the latent Gaussian space, which may not hold in highly heterogeneous clutter conditions. Future work will focus on relaxing this assumption to better capture diverse target-induced perturbations and further improve robustness.

\bibliographystyle{IEEEtran}
\bibliography{IEEEexample}

@INPROCEEDINGS{VAEalexis,
  author={Rouzoumka, Y. A. and Terreaux, E. and Morisseau, C. and Ovarlez, J.-P. and Ren, C.},
  booktitle={ICASSP 2025 - 2025 IEEE International Conference on Acoustics, Speech and Signal Processing (ICASSP)}, 
  title={Out-of-Distribution Radar Detection in Compound Clutter and Thermal Noise through Variational Autoencoders}, 
  year={2025},
  volume={},
  number={},
  pages={1-5},
  doi={10.1109/ICASSP49660.2025.10889884}}

@article{fuhrmann1992cfar,
  title={A {CFAR} adaptive matched filter detector},
  author={Fuhrmann, Daniel R and Kelly, Edward J and Nitzberg, Ramon},
  journal={IEEE Transactions on Aerospace and Electronic Systems},
  volume={28},
  number={1},
  pages={208--216},
  year={1992}
}

@article{kelly2007adaptive,
  title={An adaptive detection algorithm},
  author={Kelly, Edward J},
  journal={IEEE Transactions on Aerospace and Electronic Systems},
  number={2},
  pages={115--127},
  year={2007},
  publisher={IEEE}
}

@article{scharf2002matched,
  title={Matched subspace detectors},
  author={Scharf, Louis L and Friedlander, Benjamin},
  journal={IEEE Transactions on Signal Processing},
  volume={42},
  number={8},
  pages={2146--2157},
  year={2002},
  publisher={IEEE}
}

@inproceedings{rouzoumka2025complex,
  title={Complex-Valued Variational Autoencoders for Radar Detection in Joint Compound Gaussian Clutter and Thermal Noise},
  author={Rouzoumka, Yadang Alexis and Terreaux, Eug{\'e}nie and Morisseau, Christ{\`e}le and Ovarlez, J-P and Ren, Chengfang},
  booktitle={33rd European Signal Processing Conference},
  pages={2512--2516},
  year={2025},
  organization={Elsevier}
}

@article{zhang2024ifcmd,
  title={{IfCMD}: a novel method for radar target detection under complex clutter backgrounds},
  author={Zhang, Chenxi and Xu, Yishi and Chen, Wenchao and Chen, Bo and Gao, Chang and Liu, Hongwei},
  journal={Remote Sensing},
  volume={16},
  number={12},
  pages={2199},
  year={2024},
  publisher={MDPI}
}

@article{tyler1987,
	author = "Tyler, D. E.",
	doi = "10.1214/aos/1176350263",
	fjournal = "Annals of Statistics",
	journal = "Ann. Statist.",
	month = "03",
	number = "1",
	pages = "234--251",
	publisher = "The Institute of Mathematical Statistics",
	title = "A Distribution-Free {M}-Estimator of Multivariate Scatter",
	volume = "15",
	year = "1987"
}

@book{goodfellow2016deep,
  title={Deep learning},
  author={Goodfellow, Ian and Bengio, Yoshua and Courville, Aaron and Bengio, Yoshua},
  volume={1},
  number={2},
  year={2016},
  publisher={MIT Press Cambridge}
}

@article{liu2022flow,
  title={Flow straight and fast: Learning to generate and transfer data with rectified flow},
  author={Liu, Xingchao and Gong, Chengyue and Liu, Qiang},
  journal={arXiv preprint arXiv:2209.03003},
  year={2022}
}

@book{richards2005fundamentals,
  title={Fundamentals of radar signal processing},
  author={Richards, Mark A and others},
  volume={1},
  year={2005},
  publisher={McGraw-Hill New York}
}

@book{de2016modern,
	Editor = {Greco, M. S.  and De Maio, A.},
	Month = {Jan},
	Publisher = {SciTech Publishing},
	Title = {Modern Radar Detection Theory},
	Year = {2016}}

@article{kraut2001adaptive,
  title={Adaptive subspace detectors},
  author={Kraut, Shawn and Scharf, Louis L and McWhorter, L Todd},
  journal={IEEE Transactions on Signal Processing},
  volume={49},
  number={1},
  pages={1--16},
  year={2001},
  publisher={IEEE}
}

@article{li2025flow,
  title={Flow matching meets biology and life science: a survey},
  author={Li, Zihao and Zeng, Zhichen and Lin, Xiao and Fang, Feihao and Qu, Yanru and Xu, Zhe and Liu, Zhining and Ning, Xuying and Wei, Tianxin and Liu, Ge and others},
  journal={arXiv preprint arXiv:2507.17731},
  year={2025}
}

@article{mottier2024deinterleaving,
  title={Deinterleaving RADAR emitters with optimal transport distances},
  author={Mottier, Manon and Chardon, Gilles and Pascal, Fr{\'e}d{\'e}ric},
  journal={IEEE Transactions on Aerospace and Electronic Systems},
  volume={60},
  number={3},
  pages={3639--3651},
  year={2024},
  publisher={IEEE}
}

@article{svdd,
author = {Tax, David and Duin, Robert},
year = {2004},
month = {01},
pages = {45-66},
title = {Support Vector Data Description},
volume = {54},
journal = {Machine Learning}
}

@inproceedings{yavuz2021radar,
  title={Radar target detection with {CNN}},
  author={Yavuz, Faruk},
  booktitle={29th European Signal Processing Conference (EUSIPCO)},
  pages={1581--1585},
  year={2021},
  organization={Elsevier}
}

@inproceedings{sehgal2019automatic,
  title={Automatic radar target identification using radar cross section fluctuations and recurrent neural networks},
  author={Sehgal, Bharat and Shekhawat, Hanumant Singh and Jana, Sumit Kumar},
  booktitle={TENCON 2019-2019 IEEE Region 10 Conference (TENCON)},
  pages={2490--2495},
  year={2019},
  organization={IEEE}
}

@article{zhao2024gan,
  title={{GAN--CNN}-based moving target detector for airborne radar systems},
  author={Zhao, Yangguang and Sun, Taohan and Zhang, Jiawei and Gao, Meiguo},
  journal={IEEE Sensors Journal},
  volume={24},
  number={13},
  pages={21614--21627},
  year={2024},
  publisher={IEEE}
}

@inproceedings{svdd_detection,
  title     = {Support vector data description for radar target detection},
  author    = {Pinsolle, Jean and Rouzoumka, Yadang Alexis and Ren, Chengfang and Morisseau, Christelle and Ovarlez, Jean-Philippe},
  booktitle = {Proc. of the IEEE International Conference on Acoustics, Speech, and Signal Processing (ICASSP)},
  year      = {2026},
  note      = {to appear},
  organization = {IEEE}
}

\end{document}